\documentclass[letterpaper]{article} 
\usepackage{aaai24}  
\usepackage{times}  
\usepackage{helvet}  
\usepackage{courier}  
\usepackage[hyphens]{url}  
\usepackage{graphicx} 
\usepackage{natbib}  
\usepackage{caption} 
\usepackage{algorithm}
\usepackage{algorithmic}
\usepackage{amsmath}
\usepackage{booktabs}
\usepackage{multirow}
\usepackage{tabularray}
\usepackage{graphicx} 
\usepackage{amsfonts}
\usepackage{makecell}
\usepackage{newfloat}
\usepackage{listings}
\DeclareCaptionStyle{ruled}{labelfont=normalfont,labelsep=colon,strut=off} 
\floatstyle{ruled}
\newfloat{listing}{tb}{lst}{}
\floatname{listing}{Listing}
\title{WMFormer++: Nested Transformer for Visible Watermark Removal 
\\via Implict Joint Learning}
\author {
    Dongjian Huo\textsuperscript{\rm 1,2},
    Zehong Zhang\textsuperscript{\rm 1,2},
    Hanjing Su\textsuperscript{\rm 4},
    Guanbin Li\textsuperscript{\rm 5},
    Chaowei Fang\textsuperscript{\rm 6},
    Qingyao Wu\textsuperscript{\rm 1,3\thanks{Corresponding Author}}
}
\affiliations {
    \textsuperscript{\rm 1} School of Software and Engineering, South China University of Technology\\
    \textsuperscript{\rm 2} Key Laboratory of Big Data and Intelligent Robot, Ministry of Education\\
    \textsuperscript{\rm 3} Peng Cheng Laboratory, China\\
    \textsuperscript{\rm 4} Tencent Wechat Department\\
    \textsuperscript{\rm 5} School of Computer Science and Engineering, Sun Yat-sen University\\
    \textsuperscript{\rm 6} Xidian University\\
    huodongjian0603@gmail.com
}

\usepackage{bibentry}

\begin{document}

\maketitle

\begin{abstract}
Watermarking serves as a widely adopted approach to safeguard media copyright. In parallel, the research focus has extended to watermark removal techniques, offering an adversarial means to enhance watermark robustness and foster advancements in the watermarking field. 
Existing watermark removal methods mainly rely on UNet with task-specific decoder branches——one for watermark localization and the other for background image restoration. 
However, watermark localization and background restoration are not isolated tasks; precise watermark localization inherently implies regions necessitating restoration, and the background restoration process contributes to more accurate watermark localization.
To holistically integrate information from both branches, we introduce an implicit joint learning paradigm. This empowers the network to autonomously navigate the flow of information between implicit branches through a gate mechanism. 
Furthermore, we employ cross-channel attention to facilitate local detail restoration and holistic structural comprehension, while harnessing nested structures to integrate multi-scale information.
Extensive experiments are conducted on various challenging benchmarks to validate the effectiveness of our proposed method. The results demonstrate our approach's remarkable superiority, surpassing existing state-of-the-art methods by a large margin.
\end{abstract}

\section{Introduction}
In today's digital era, images have become the predominant means of recording and conveying information, often adorned with visible watermarks to assert copyright or ownership. While overlaying visible watermarks is an efficient defense against unauthorized use, it also raises concerns about the robustness of such watermarks in the face of modern watermark removal techniques. 
This has propelled the emergence of watermark removal as a critical research area, dedicated to assessing and enhancing the resilience of visible watermarks through strategic adversarial approaches.
This paper delves into the challenges and advancements in watermark removal, with a specific focus on enhancing the effectiveness of visible watermarks in safeguarding digital media.

\begin{figure}[t]
\centering
\includegraphics[width=\linewidth]{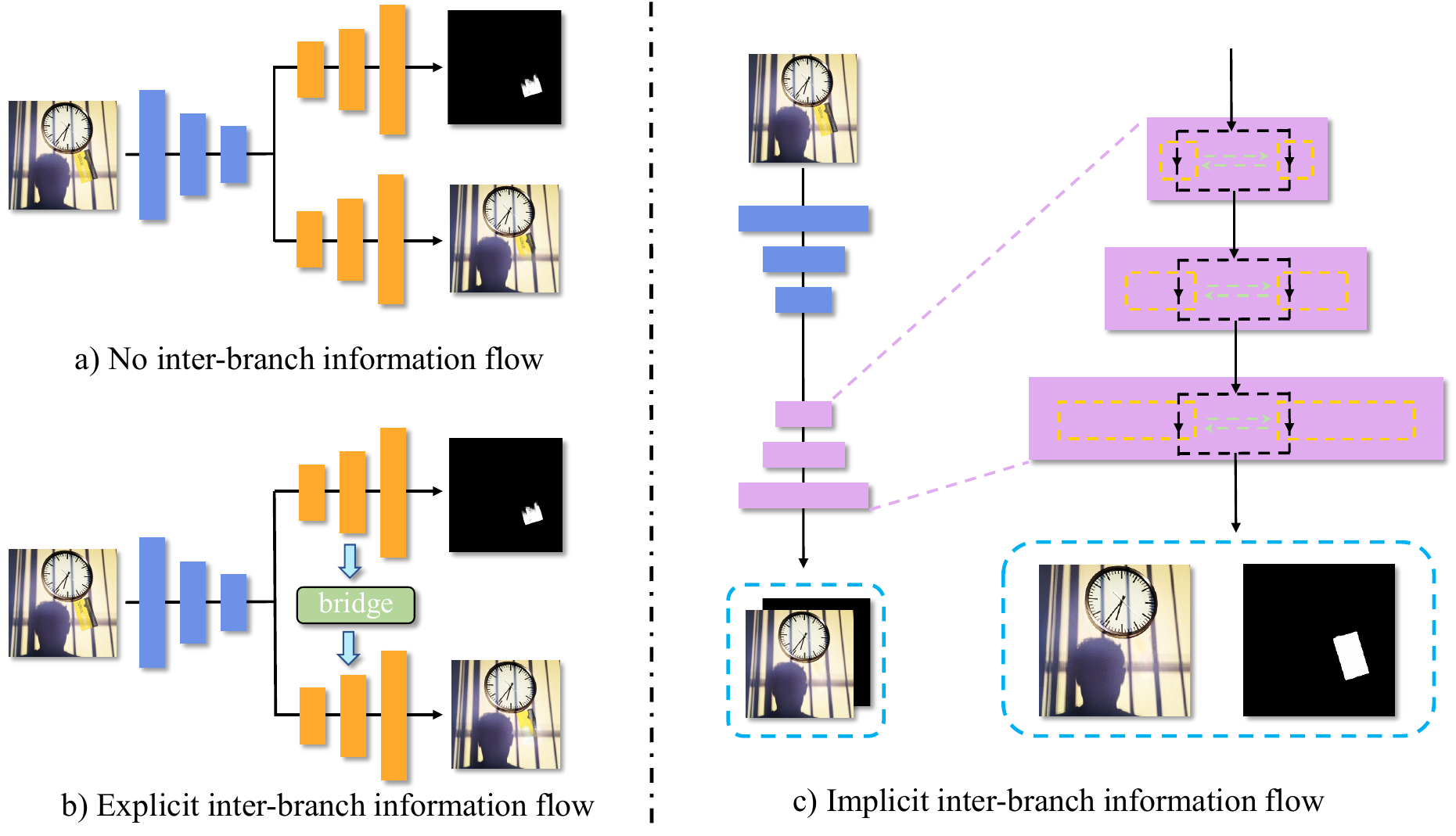} 


\caption{
Compared with the conventional independent multi-branch or hand-crafted bridge methods, we realize the joint collaboration of background restoration and watermark localization in an implicit way.
}
\label{figure1}
\vspace{-12pt}
\end{figure}

Amidst the rapid evolution of deep learning, data-driven methodologies have surged to the forefront in the realm of watermark removal, achieving remarkable strides.
~\citet{hertz2019blind} introduced a multi-task network for blind watermark removal from single images, achieving commendable results through multiple decoder branches dedicated to distinct tasks, including watermark location and background restoration.
~\citet{cun2021split} proposed a novel two-stage framework with stacked attention-guided ResUNets to address texture non-harmony challenges in watermark removal.
Following Hertz and Cun's pioneering introduction of multi-decoder multi-stage networks for watermark removal, subsequent studies~\cite{liang2021visible,zhao2022visible,sun2023denet} have predominantly focused on refining modules and enhancing coarse-level aspects within this paradigm.



However, this paradigm treats watermark localization and image restoration tasks as isolated processes within separate decoder branches (as depicted in Fig.~\ref{figure1}(a)), potentially underutilizing the inter-branch information exchange. Recognizing this, ~\citet{liang2021visible} meticulously designed modules to control information flow between watermark and background branches, enhancing background restoration. The effectiveness of this module implies that the background branch contains numerous watermark fragments, which, in turn, can be leveraged to enhance watermark localization. While a viable approach would be to design a handcrafted module to guide inter-branch information flow, this proves challenging (as shown in Fig.~\ref{figure1}(b)). Instead of relying on such intricate handcrafted designs, we propose a Transformer-based method that treats the two interdependent branches as a unified entity, utilizing Gated-Dconv Feed-Forward Network~\cite{zamir2022restormer} as a distributor to effectively guide the flow of information (as illustrated in Fig.~\ref{figure1}(c)). Furthermore, drawing inspiration from the implicit modeling of spatial relationships among adjacent pixels and the global context of all pixels in~\cite{zamir2022restormer}, we exploit the Cross-channel Multi-head Attention mechanism. This mechanism enhances both the quality of localized background restoration and the accuracy of global watermark localization. To accommodate the variance in understanding the watermark's localized restoration area across different scales, we introduce a nested network design, facilitating the fusion of structural information from various scales.
We hope that this work will inspire deeper contemplation within the watermark removal framework, infusing renewed vitality into the field and paving the way for innovative advancements in watermark removal techniques.
Extensive experiments on various challenging benchmarks, including LOGO-H, LOGO-L, LOGO-Gray~\cite{cun2021split}, and CLWD~\cite{liu2021wdnet}, along with qualitative intermediate visualizations, validate the effectiveness of our proposed method. 

Our contributions can be summarized as follows:
\begin{itemize}
\item We proposed a novel Transformer-based network that utilizes a single decoder to handle both watermark localization and background restoration tasks concurrently, eliminating the requirement for additional refinement steps.
\item We employ the Gated-Dconv Feed-Forward Network for effective information flow control, while the Cross-channel Multi-head Attention ensures detailed local reconstruction and comprehensive structural understanding. Additionally, we harness the power of a nested network design to enhance the comprehension of restoration areas across different scales.
\item Through extensive experimental evaluations on different datasets, we demonstrated the superiority and effectiveness of our proposed method, achieving new state-of-the-art performance and producing high-quality output.
\end{itemize}

\section{Related Work}
\subsection{Watermark Removal}
In the realm of digital copyright protection, digital watermarks have assumed a critical role. Early approaches~\cite{pei2006novel,park2012identigram} for watermark removal heavily relied on manually crafted features and necessitated user intervention for watermark localization, leading to usability challenges. Some researchers~\cite{dekel2017effectiveness,gandelsman2019double} explored methods based on multiple images, but these demanded extensive prior knowledge and were limited in their applicability to specific samples.

Recent advancements in deep learning-based techniques have shown significant promise in various computer vision tasks, including visible watermark removal. 
~\citet{hertz2019blind} pioneered a groundbreaking approach by blindly removing visual motifs from images and introducing the single encoder with multi-decoder architecture for multi-task watermark removal, achieving impressive performance. ~\citet{cun2021split} further enhanced watermark removal performance by proposing two-stage networks for prediction and refinement, solidifying the multi-decoder multi-stage framework as a mainstream solution for watermark removal. Building on this backdrop, ~\citet{liang2021visible} introduced a set of intricate modules to enhance the quality of generated images, while ~\citet{sun2023denet} employed a contrastive learning strategy to disentangle high-level embedding semantic information of images and watermarks.

While the multi-decoder multi-stage framework has been instrumental, it is crucial to reassess its rationale and adopt a more concise architecture. 
By seeking a more streamlined solution, we can further advance research in the field of watermark removal, providing more practical and sustainable approaches for real-world applications.

\subsection{Image Restoration}
Watermark removal shares a resemblance to image dehazing~\cite{he2010single}, deraining~\cite{ren2019progressive}, and shadow removal~\cite{cun2020towards} tasks in image restoration. They all involve recovering the source image from a damaged version, but specific differences exist that cannot be ignored in practical applications.

In dehazing and deraining, the interference factors (haze and raindrops) permeate the entire image, with repeated patterns and semantics present within and across images. In contrast, watermarks are usually localized in specific areas of the image, with each watermark representing unique information independently. Moreover, shadow removal deals with meaningless grey areas, while watermarks symbolize media copyright and typically exhibit meaningful and colorful content. These distinctions make watermark removal a distinct and challenging research domain.

Recently, the adoption of a novel Transformer architecture~\cite{zamir2022restormer} in image restoration has shown remarkable results, inspiring us to explore the application of similar modules in the field of watermark removal.

\subsection{UNet}
UNet~\cite{ronneberger2015u} is a classic neural network known for its effectiveness in image semantic segmentation. Its U-shaped architecture with symmetric encoding and decoding pathways enables it to capture both global and local features, achieving remarkable performance in image segmentation tasks.

UNet++~\cite{zhou2018unet++} is an improved version of UNet. It builds upon the original UNet by introducing multi-level encoding-decoding paths. This design allows the network to progressively fuse feature information across layers, resulting in a better grasp of semantic information at different scales and further enhancing segmentation performance.

While UNet has been widely applied as a foundational framework in the field of watermark removal, the application of the UNet++ architecture in watermark removal remains relatively unexplored and lesser-known.

\begin{figure*}[t]
\centering
\includegraphics[width=0.9\textwidth]{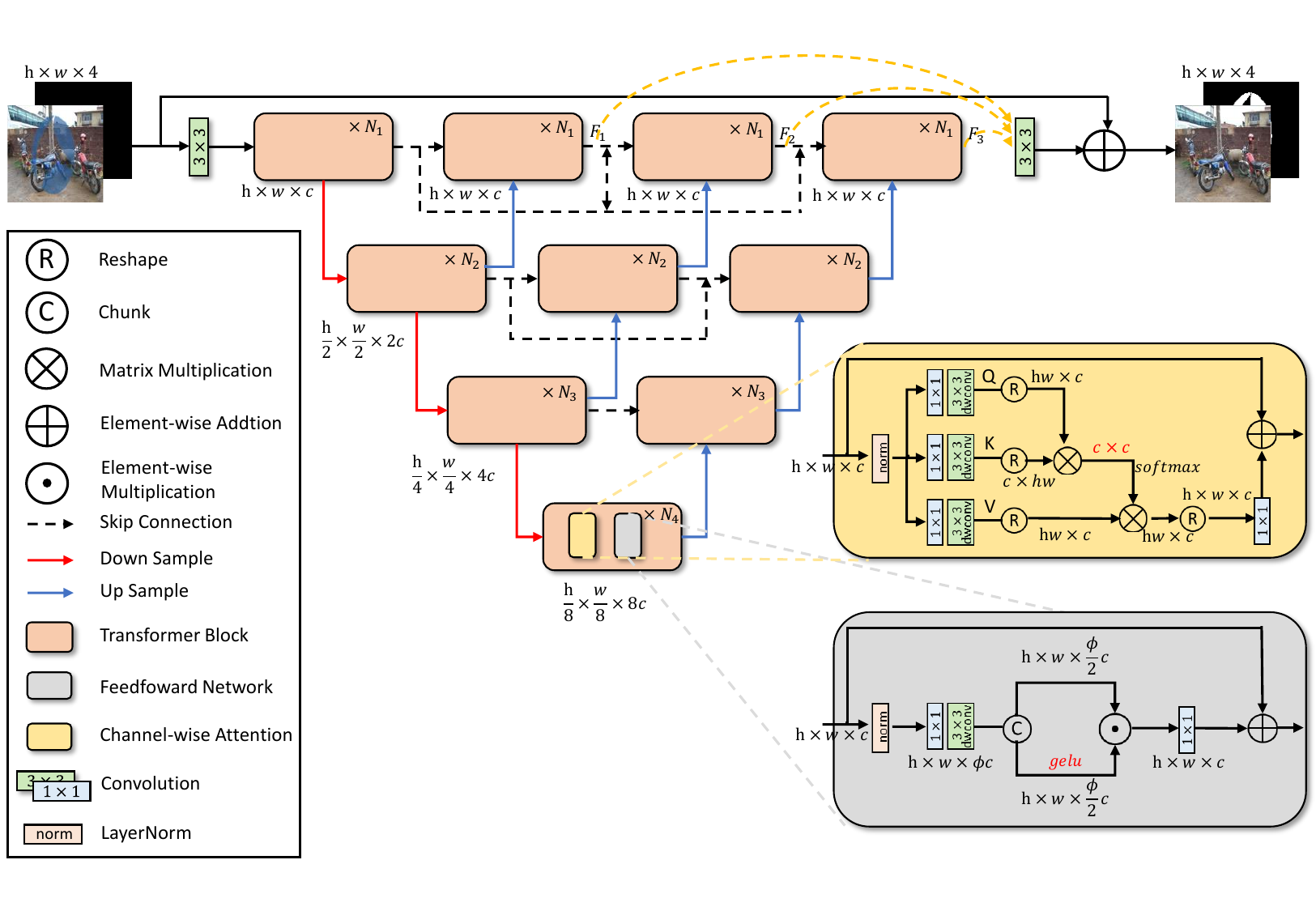} 
\caption{The overview of our proposed WMFormer++. In the training phase, given a watermarked image, it is first concatenated with a all-black grayscale image. The concatenated input is then fed into an embedding layer, which produces the embedded feature. This embedded feature is further processed through a nested stack of transformer blocks, enabling both encoding and decoding steps to yield the output feature. Along different pathways of the embedded feature, distinct hierarchical output features are obtained at the topmost decoder. Each hierarchical output feature passes through a shared prediction head, generating predictions for watermark-free images and watermark masks, with supervision from the ground truth watermark-free image and watermark mask.
In the testing phase, a watermarked image traverses through the complete network to obtain the final prediction. Specifically, only the output feature from the final Transformer blocks ($F_3$) is utilized for watermark-free image reconstruction.}
\label{figure2}
\end{figure*}
\section{Methodology}

In this paper, we have reimagined the development of the watermark removal domain from a framework perspective. Our goal is to challenge the current popular paradigm of multi-decoder removal networks and design a concise watermark removal framework while exploring the potential of transformers in watermark removal tasks. As a result, we propose a simple yet effective Transformer-based implicit joint training framework called WMFormer++.

\subsection{Overall Architecture}
For a given watermarked image $J \in \mathbb{R}^{H\times W\times 3}$, our network produces a watermark-free restored background image $\hat{I} \in \mathbb{R}^{H\times W\times 3}$ as output, with the grayscale watermark mask $\hat{M} \in \mathbb{R}^{H\times W\times 1}$ being generated as a byproduct. 
The network architecture consists of a multi-stage encoder and a nested decoder, with both components composed of fundamental Transformer blocks.

\subsection{Basic Module}
A novel type of Transformer~\cite{zamir2022restormer} serves as the basic building block of the entire network, composed of two key components: Cross-channel Multi-head Attention and Gated-Dconv Feed-Forward Network.
\\ \hspace*{\fill} \\
\noindent \textbf{(i) Cross-channel Multi-head Attention.} 
This component serves to boost local restoration quality while enabling comprehensive restoration region anchoring.
Initially, the module utilizes $1\times 1$ convolutions and $3\times 3$ depth-wise convolutions to gather spatial information from neighboring pixels. 
Subsequently, it calculates cross-channel attention maps to implicitly establish global context semantics across pixels. 

Given the input $X$, the attention for the i-th head is computed using the following formula: 
\begin{align}
Q_i, K_i, V_i &= XW_i^Q, XW_i^K, XW_i^V, \\
head_i &= V_i \text{ softmax}(\sigma K_i^TQ_i),
\end{align}
where $Q_i, K_i$, and $V_i$ represent the query, key, and value for each head $i$, respectively. $W_i^Q, W_i^K,$ and $W_i^V$ are three projection matrices. $\sigma$ is a learnable parameter.
\\ \hspace*{\fill} \\
\noindent \textbf{(ii) Gated-Dconv Feed-Forward Network.} 
This module utilizes a gating mechanism to enhance information flow within the network, facilitating the collaborative optimization of the two implicit branches: watermark localization and background restoration. 
This synergistic approach enables them to collectively refine their respective tasks.

The entire process can be represented using the following formula:
\begin{align}
X_1, X_2 &= split\_half\left(XW_1, axis=channel\right), \\
Y &= \left(gelu\left(X_1\right)\odot X_2\right)W_2 + X,
\end{align}
where $W_1$ and $W_2$ correspond to the projection matrices responsible for augmenting and reducing the channel dimensions of $X$. The output of the module is denoted by $Y$.

\subsection{Encoder}
The Encoder module follows a similar structure as the UNet~\cite{ronneberger2015u}, comprising four hierarchical levels. As data progresses through each level, the feature map dimensions remain unchanged, but its channel count doubles and size is halved before transitioning to the subsequent level. This progressive encoding leads to the compression of the original input, resulting in a latent code enriched with semantic information. 

In general, the latent code generated by the final level acts as input for the Decoder. Notably, in contrast to the conventional approach, we harness the latent codes generated at each level during the decoding process. This synergistic interaction among nested decoders empowers the creation of UNet models with varying depths, allowing for a more adaptable and flexible architecture.

\subsection{Decoder}
The latent code encapsulates two crucial aspects of information: the overall structure of the repair regions, including the position of the watermark within the image, and various reconstruction details. The latent codes from different levels of the encoder harbor distinct information due to variations in encoding depth. To capitalize on this diversity, we decode latent codes from different levels to construct independent UNet pathways of varying depths. These pathways are then interlinked through dense connections, creating a nested UNet architecture that enhances feature fusion. The integration of UNet pathways at different depths enables the network to precisely locate watermark regions. Simultaneously, through skip connections at the same level, distinct UNet pathways can learn specific reconstruction details to aid in background restoration.

For a specific pathway of the decoder, the decoding process mirrors the reverse of the encoding procedure. For each level's output, the channel count is reduced by half while the size doubles before moving to the previous level. This stepwise decoding gradually restores the latent code back to enriched feature maps containing discrete watermark and background information. Finally, through the last prediction head, both watermark-free images and watermark masks are generated. 

At each decoder level, a decision point arises regarding the output strategy. The network can either upsample the feature by a factor of two while reducing the number of channels, thereby progressing to the previous decoder level, or it can directly transmit the feature to the decoder at the same level. The network benefits from these intra-level skip connections due to the distinct semantic information encoded within features originating from different depths of latent codes, even when they are at the same decoder level.

\subsection{Objective Function}
In our proposed framework, we leverage the nested architecture as the backbone, which can be seen as an integration of UNets with varying depths. To ensure that each UNet is capable of both image restoration and watermark extraction, during the training phase, we allow the refined features from each UNet's output to pass through a shared prediction head. This step facilitates the generation of corresponding watermark mask $\hat{M}^d$ and background image $\hat{I}^d$, which are then supervised using the ground truth watermark mask $M$ and background image $I$.

Consistent with previous works~\cite{hertz2019blind}, we employ binary cross-entropy loss to ensure the close alignment between the predicted watermark mask $\hat{M}^d$ and the ground-truth watermark mask $M$. 


\begin{align}
\mathcal{L}_{\text {mask}}^d=-\sum_{i, j}(&\left(1-M_{i,j}\right) \log (1-\hat{M}_{i, j}^d) 
 \notag \\
&+M_{i, j} \log \hat{M}_{i, j}^d) . 
\end{align}

As for the reconstruction task, we utilize the $L_1$ loss to ensure that the refined watermark-free image closely approximates the ground-truth image.
\begin{equation}
\mathcal{L}_{\text {image}}^d=\|I-\hat{I}^d\|_1.
\end{equation}

Moreover, inspired by the works of ~\cite{cun2021split,liang2021visible}, we incorporate an additional deep perceptual loss~\cite{johnson2016perceptual} to enhance the output's overall quality.
\begin{equation}
\mathcal{L}_{\text {perc}}^d=\sum_{k \in 1,2,3}\left\|\Phi_{vgg}^k(\hat{I}^d)-\Phi_{vgg}^k(I)\right\|_1,
\end{equation}
where $\Phi_{vgg}^k(\cdot)$ denotes the activation map of k-th layer in the pre-trained VGG16~\cite{simonyan2014very}.

In conclusion, we aggregate all the aforementioned loss functions and introduce controllable hyper-parameters to form the final loss function. The losses from different depths of the UNet are combined harmoniously, playing a collective role in shaping the overall loss.
\begin{equation}
\label{eq_loss}
\mathcal{L}_{\text {all}}= \sum_{d \in 1,2,3} \lambda^d\left( \mathcal{L}_{\text {image}}^d+\lambda_{\text {mask}} \mathcal{L}_{\text {mask}}^d+\lambda_{\text {perc}} \mathcal{L}_{\text {perc}}^d\right).
\end{equation}

\begin{table}[!t]
\renewcommand\arraystretch{1.3}
	\begin{center}
		\scalebox{1.0}{
			\begin{tabular}{ccccc}
				\toprule  
				\toprule  
				UNet  & Transformer & Nested & deep\_sup  & PSNR \\
				\midrule  
				$\checkmark$&  & & & 34.87 \\
				$\checkmark$&  $\checkmark$& & & 46.09 \\
				$\checkmark$&  $\checkmark$&$\checkmark$ & &  46.35 \\
				$\checkmark$&  $\checkmark$&$\checkmark$& $\checkmark$ &  \textbf{47.05} \\
				
				\bottomrule 
		\end{tabular}}
	\end{center}\caption{Ablation study on the LOGO-L dataset.}
\label{table1}
\end{table}

\begin{figure}[t]
\centering
\includegraphics[width=\linewidth]{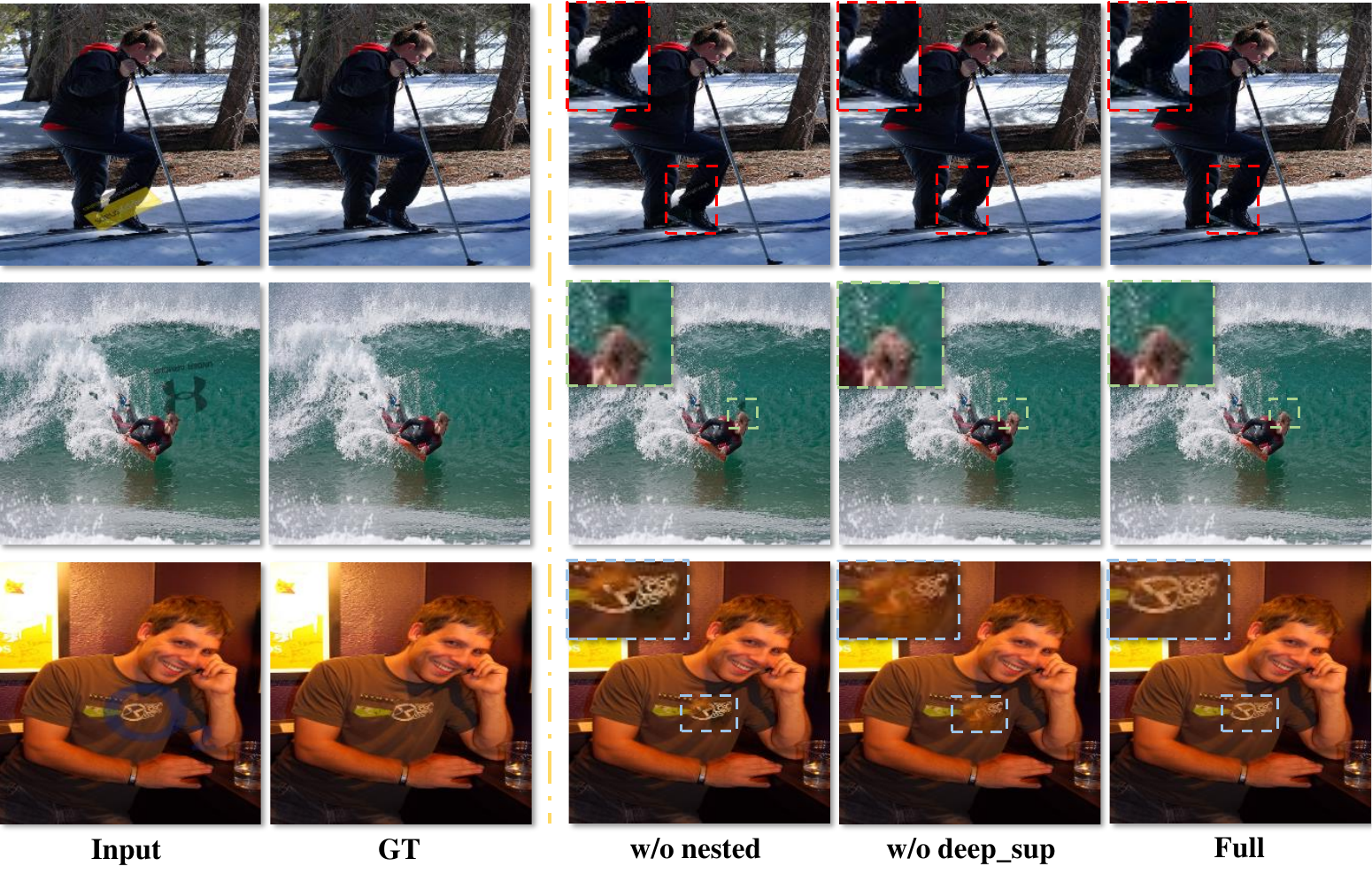} 
\caption{The impact of different modules on the final result's visual quality.}
\label{figure3}
\end{figure}

\section{Experiment}
This section begins with an introduction to the datasets used and outlines the implementation details. Subsequently, we conduct comprehensive ablation experiments to thoroughly investigate the influences of different architectural designs. Finally, we evaluate the performance of our WMFormer++ against state-of-the-art methods on different datasets, including LOGO-L, LOGO-H, LOGO-Gray~\cite{cun2021split}, and CLWD~\cite{liu2021wdnet}. The experimental findings demonstrate the effectiveness of our approach both qualitatively and quantitatively.

\begin{table}[t]
\renewcommand\arraystretch{1.3}
	\begin{center}
		\scalebox{1.0}{
			\begin{tabular}{p{2cm}<{\centering}|p{1.5cm}<{\centering}p{1.5cm}<{\centering}}
				\toprule  
				\toprule  
				Method  & F1 & IoU(\%) \\
				\midrule  
				BVMR & 0.7871 & 70.21 \\
				WDNet & 0.7240 & 61.20 \\
                    SplitNet & 0.8027 & 71.96 \\
                    SLBR & 0.8234 & 74.63 \\
                    \hline
                    Ours($F_2$) & 0.8619 & 79.04 \\
                    Ours($F_3$) & \textbf{0.8769} & \textbf{81.38} \\
				\bottomrule 
		\end{tabular}}
	\end{center}\caption{Quantitative evaluation of watermark masks predicted by different methods on CLWD dataset. Among them, "Ours($F_2$)" and "Ours($F_3$)" represent the watermark masks generated from the corresponding refined feature maps $F_2$, $F_3$.}
\label{table2}
\end{table}

\begin{table*}
\centering
\resizebox{\linewidth}{!}{
\renewcommand\arraystretch{1.5}
\begin{tabular}{c|c| c|c|c| c|c|c| c|c|c} 
\toprule[1pt]\hline
\multicolumn{2}{c|}{\multirow{2}{*}{Methods}} &                   \multicolumn{3}{c|}{LOGO-H} & \multicolumn{3}{c|}{LOGO-L} & \multicolumn{3}{c}{LOGO-Gray}      \\ 
\cline{3-11}
\multicolumn{2}{c|}{}                         & PSNR $\uparrow$ & SSIM $\uparrow$  & LIPIS $\downarrow$     & PSNR $\uparrow$ & SSIM $\uparrow$  & LIPIS $\downarrow$     & PSNR $\uparrow$ & SSIM $\uparrow$  & LIPIS $\downarrow$                 \\ 
\hline
\multirow{4}{*}{MDMS} & BVMR                  & 36.51 & 0.9799 & 2.37       & 40.24 & 0.9895 & 1.26       & 38.90 & 0.9873 & 1.15                  \\
                      & SplitNet              & 40.05 & 0.9897 & 1.15       & 42.53 & 0.9924 & 0.87       & 42.01 & 0.9928 & 0.73                    \\
                      & SLBR                  & 40.56 & 0.9913 & 1.06       & 44.10 & 0.9947 & 0.70       & 42.21 & 0.9936 & 0.69                      \\
                      & DENet                 & 40.83 & 0.9919 & 0.89       & 44.24 & 0.9954 & 0.54       & 42.60 & 0.9944 & 0.53                       \\ 
\hline
\multirow{5}{*}{SDSS} & UNet                  & 30.51 & 0.9612 & 5.44       & 34.87 & 0.9814 & 2.97       & 32.15 & 0.9728 & 3.53             \\
                      & SIRF                  & 32.35 & 0.9673 & 8.01       & 36.25 & 0.9825 & 6.55       & 34.33 & 0.9782 & 6.72                      \\
                      & BS$^2$AM                  & 31.93 & 0.9677 & 4.45       & 36.11 & 0.9839 & 2.23       & 32.91 & 0.9754 & 3.05                      \\ 
                      & DHAN                  & 35.68 & 0.9809 & 6.61       & 38.54 & 0.9887 & 5.91       & 36.39 & 0.9836 & 5.94                     \\
\cline{2-11}
                      & \textbf{WMFormer++}            & \textbf{44.64}     & \textbf{0.9950}      & \textbf{0.50}          & \textbf{47.05} & \textbf{0.9970} & \textbf{0.31}          & \textbf{46.29}     & \textbf{0.9970}      & \textbf{0.21}            \\
\bottomrule[1pt]
\end{tabular}
}

\caption{Quantitative comparisons of the proposed WMFormer++ with other state-of-the-art methods on the LOGO series datasets. The best results are highlighted in bold. Here, MDMS denotes methods with multiple decoder branches or multi-stage refinement networks, while SDSS represents methods without multiple decoder branches or multi-stage designs.}
\label{table3}
\end{table*}

\begin{table}
\centering

\renewcommand\arraystretch{1.5}
\scalebox{0.8}{
\begin{tabular}{c| c|c|c|c} 
\toprule\hline
Method & PSNR $\uparrow$ & SSIM $\uparrow$  & RMSE $\downarrow$ & RMSEw $\downarrow$ \\
\midrule
UNet & 23.21 & 0.8567 & 19.35 & 48.43 \\
DHAN& 35.29 & 0.9712 & 5.28  & 18.25   \\
BVMR & 35.89 & 0.9734 & 5.02  & 18.71 \\
SplitNet & 37.41 & 0.9787 & 4.23  & 15.25  \\
SLBR & 38.28 & 0.9814 & 3.76  & 14.07  \\
\midrule
WMFormer++ & \textbf{39.36} & \textbf{0.9830} & \textbf{3.25} & \textbf{11.47} \\
\bottomrule
\end{tabular}}
\caption{Quantitative comparisons of the proposed WMFormer++ with other state-of-the-art methods on CLWD dataset. The best results are highlighted in bold.}
\label{table4}
\end{table}

\subsection{Dataset}
This paper leverages various datasets, namely LOGO-L, LOGO-H, LOGO-Gray, and CLWD, for conducting the experiments.

The \textbf{LOGO series} dataset contains varying watermarks, such as LOGO-L with 12,151 training and 2,025 testing images, LOGO-H with similar quantities but larger watermark sizes and transparency, and LOGO-Gray with grayscale watermarks. 

\textbf{CLWD} dataset contains 60,000 watermarked images, with 160 colored watermarks for training and 10,000 images with 40 colored watermarks for testing. The watermarks are collected from open-sourced logo images websites. The original images in the training and test sets are randomly chosen from the PASCAL VOC 2012~\cite{everingham2015pascal} dataset. The transparency of the watermarks in CLWD is set in the range of 0.3 to 0.7, and the size, locations, rotation angle, and transparency of each watermark are randomly set in different images.

\subsection{Implementation Detail}
We implemented our method using Pytorch~\cite{paszke2019pytorch} and conducted experiments on the aforementioned datasets. For training, we set the input image size as $256\times256$. The AdamW~\cite{loshchilov2017decoupled} optimizer was chosen with an initial learning rate of 3e-4, a batch size of 8, and momentum parameters $\beta_1=0.9$ and $\beta_2=0.999$. The hyper-parameters $\lambda^d$, $\lambda_{mask}$, and $\lambda_{vgg}$ in Eqn.(\ref{eq_loss}) were empirically set as 1, 1, and 0.25, respectively, after several trials by observing the quality of predicted masks and reconstructed images.

To evaluate the effectiveness of our method, we employ widely recognized metrics in the field. Specifically, on the LOGO series datasets, we use Peak Signal-to-Noise Ratio (PSNR), Structural Similarity (SSIM)~\cite{wang2004image}, and the deep perceptual similarity metric (LPIPS)~\cite{zhang2018unreasonable} for evaluation. On the CLWD dataset, we utilize PSNR, SSIM, Root Mean Square Error (RMSE), and weighted RMSE (RMSEw) as evaluation metrics.

\subsection{Ablation Study}
\textbf{Analysis of Individual Modules}: As depicted in Tab.~\ref{table1}, we conducted a comprehensive evaluation of each module's effectiveness in our framework through incremental removal and addition. Initially, we started with a basic UNet structure, as referenced ~\cite{ronneberger2015u}, with its performance displayed in the first row of the table. Subsequently, by replacing all convolutional modules in UNet's encoder and decoder with Transformer modules, we observed a notable improvement in performance, as shown in the second row. This highlights the significant potential of applying Transformers in the field of watermark removal.

Furthermore, we devised a nested Transformer model as illustrated in Fig.~\ref{figure2}, and its performance is documented in the third row. It is worth noting that, for a fair comparison, the Transformer-based UNet used in the second row employed a wider UNet to achieve a similar parameter count as the model experimented with in the third row. The comparison between the second and third rows provides compelling evidence for the efficacy of the nested architecture.

Finally, We applied the supervision signals to UNets with different depths, and the performance is presented in the fourth row. The comparison between the third and fourth rows clearly demonstrates that the model with deep supervision exhibits superior performance, achieving the best results across all experiments. This further validates the significance of incorporating deep supervision in our approach.
\\ \hspace*{\fill} \\
\noindent \textbf{Visualization of Individual Modules}: For a more intuitive presentation, we also generated corresponding visual results to complement the findings in Tab.~\ref{table1}. By zooming in on specific regions of the restored background images, we aimed to provide a clearer display. As shown in Fig.~\ref{figure3}, with the gradual completion of the full model, the quality of the images significantly improves. The step-by-step enhancement of image quality through the addition of modules serves as concrete evidence of the effectiveness of each design in our approach. These visual results further reinforce the significance of our contributions.

\begin{figure*}[!t]
\centering
\includegraphics[width=1\textwidth]{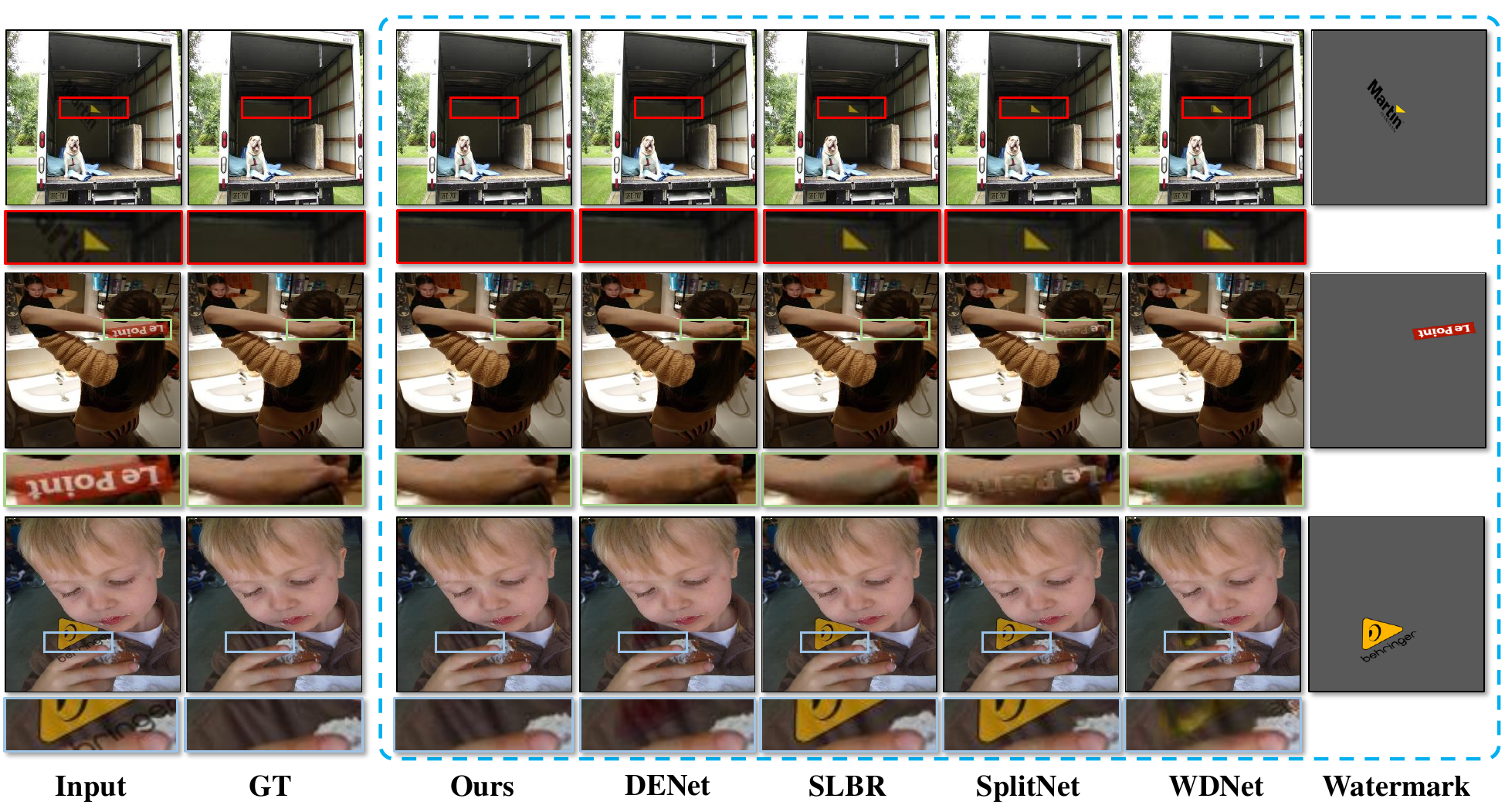} 
\caption{Qualitative comparison of background image restored by different methods on LOGO-L dataset. Our approach excels at producing superior-quality reconstructed background images, free from any visible watermark artifacts.}
\label{figure4}
\end{figure*}

\begin{figure}[!t]
\centering
\includegraphics[width=0.47\textwidth]{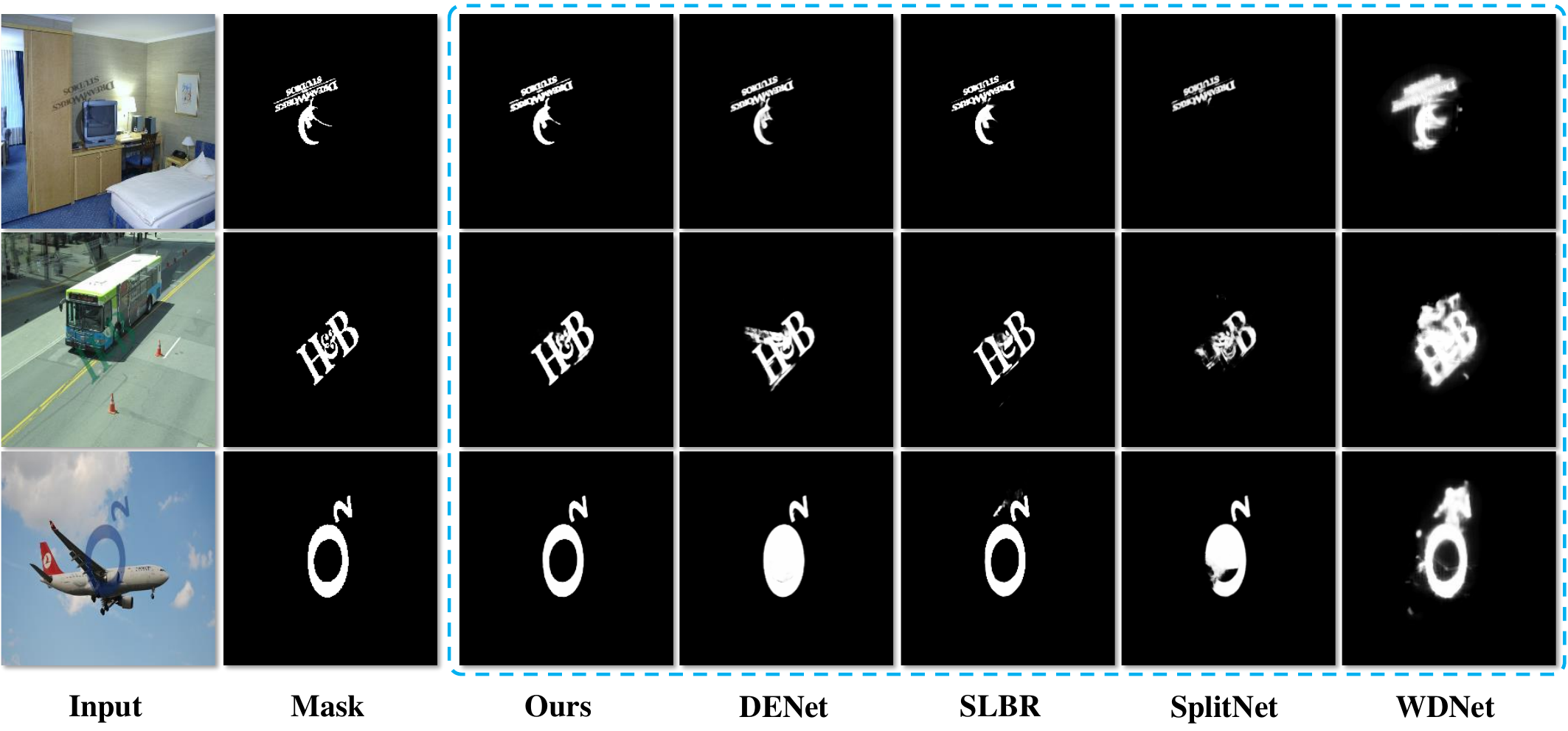} 
\caption{Qualitative comparison of watermark masks predicted by different methods on LOGO-L dataset. Our proposed WMFormer++ consistently generated watermark masks with the highest quality and accuracy.}
\label{figure5}
\vspace{-8pt}
\end{figure}

\subsection{Comparisons with State-of-the-art Methods}
\textbf{Background Restoration.} The quantitative and qualitative comparison of our proposed WMFormer++ with other existing watermark removal methods are summarized in Tab.~\ref{table3}, Tab.~\ref{table4}, and Fig.~\ref{figure4}. Among them, BVMR~\cite{hertz2019blind}, SplitNet~\cite{cun2021split}, SLBR~\cite{liang2021visible}, and DENet~\cite{sun2023denet} are the latest technologies dedicated to watermark removal and generally adopt the multi-decoder multi-stage framework. On the other hand, Unet~\cite{ronneberger2015u}, SIRF~\cite{zhang2018single}, BS$^2$AM~\cite{cun2020improving}, and DHAN~\cite{cun2020towards} are migrated from related tasks, such as blind image harmonization and shadow removal, and typically utilize a single-decoder branch for image restoration. Overall, the multi-decoder multi-stage methods tend to exhibit better performance.

In contrast, our framework follows a single-decoder approach and outperforms all the other methods on the four datasets, achieving a new state-of-the-art performance. Our results significantly surpass previous non-multi-decoder multi-stage methods and even demonstrate unprecedented superiority over multi-decoder multi-stage methods by a large margin. The comprehensive experimental results provide strong evidence for the effectiveness of our approach.

It is worth noting that our design does not rely on complex modules like those in Liang $\emph{et~al.}$~\cite{liang2021visible} and Cun $\emph{et~al.}$~\cite{cun2021split}. Instead, we use only two fundamental components, the Transformer block, and a small number of convolutional layers, for the entire model construction. This simplicity further highlights the efficiency and effectiveness of our approach in achieving outstanding watermark removal results.

\noindent \textbf{Watermark Location.} To further demonstrate the superiority of our method, we conducted both qualitative and quantitative experiments on the watermark masks, a byproduct of the watermark removal network, as shown in Tab.~\ref{table2} and Fig.~\ref{figure5}. It is evident that not only does our approach outperform previous methods in terms of numerical metrics, but the generated watermark masks are also more complete and accurate.

In Tab.~\ref{table2}, the quantitative evaluation indicates that our method achieves significantly better results compared to previous approaches in terms of F1 and IoU. Moreover, Fig.~\ref{figure5} visually presents the watermark masks generated by our method, showcasing their high quality and precision.

These results serve as compelling evidence of the effectiveness and robustness of our approach, not only in removing watermarks from images but also in generating accurate and complete watermark masks as a valuable side benefit. This highlights the holistic superiority of our method in watermark removal and its related applications.

\section{Conclusion}

In this paper, we introduced WMFormer++, a novel nested Transformer network for visible watermark removal. 
Conventional methods often struggle to fully exploit the inherent relationship between watermark localization and background restoration tasks, limited by their distinct decoder branches. Based on this observation, the key innovation of our approach lies in utilizing a single decoder branch to simultaneously handle watermark localization and background restoration tasks, surpassing previous multi-decoder methods without the need for additional refinement. Notably, the construction of our network is intentionally kept straightforward, relying exclusively on fundamental Transformer blocks. 
Moreover, we introduce a nested mechanism, which aids in enhancing feature fusion and contextual understanding across diverse scales, facilitating improved performance.
Extensive comparisons on various datasets demonstrate that WMFormer++ achieves state-of-the-art performance both qualitatively and quantitatively. Furthermore, through comprehensive ablation experiments and visualizations, we have shown the effectiveness and interpretability of our proposed method.

\bibliography{aaai24}

\begin{thebibliography}{25}
\providecommand{\natexlab}[1]{#1}

\bibitem[{Cun and Pun(2020)}]{cun2020improving}
Cun, X.; and Pun, C.-M. 2020.
\newblock Improving the harmony of the composite image by spatial-separated
  attention module.
\newblock \emph{IEEE Transactions on Image Processing}, 29: 4759--4771.

\bibitem[{Cun and Pun(2021)}]{cun2021split}
Cun, X.; and Pun, C.-M. 2021.
\newblock Split then refine: stacked attention-guided ResUNets for blind single
  image visible watermark removal.
\newblock In \emph{Proceedings of the AAAI conference on artificial
  intelligence}, volume~35, 1184--1192.

\bibitem[{Cun, Pun, and Shi(2020)}]{cun2020towards}
Cun, X.; Pun, C.-M.; and Shi, C. 2020.
\newblock Towards ghost-free shadow removal via dual hierarchical aggregation
  network and shadow matting gan.
\newblock In \emph{Proceedings of the AAAI Conference on Artificial
  Intelligence}, volume~34, 10680--10687.

\bibitem[{Dekel et~al.(2017)Dekel, Rubinstein, Liu, and
  Freeman}]{dekel2017effectiveness}
Dekel, T.; Rubinstein, M.; Liu, C.; and Freeman, W.~T. 2017.
\newblock On the effectiveness of visible watermarks.
\newblock In \emph{Proceedings of the IEEE Conference on Computer Vision and
  Pattern Recognition}, 2146--2154.

\bibitem[{Everingham et~al.(2015)Everingham, Eslami, Van~Gool, Williams, Winn,
  and Zisserman}]{everingham2015pascal}
Everingham, M.; Eslami, S.~A.; Van~Gool, L.; Williams, C.~K.; Winn, J.; and
  Zisserman, A. 2015.
\newblock The pascal visual object classes challenge: A retrospective.
\newblock \emph{International journal of computer vision}, 111: 98--136.

\bibitem[{Gandelsman, Shocher, and Irani(2019)}]{gandelsman2019double}
Gandelsman, Y.; Shocher, A.; and Irani, M. 2019.
\newblock " Double-DIP": unsupervised image decomposition via coupled
  deep-image-priors.
\newblock In \emph{Proceedings of the IEEE/CVF Conference on Computer Vision
  and Pattern Recognition}, 11026--11035.

\bibitem[{He, Sun, and Tang(2010)}]{he2010single}
He, K.; Sun, J.; and Tang, X. 2010.
\newblock Single image haze removal using dark channel prior.
\newblock \emph{IEEE transactions on pattern analysis and machine
  intelligence}, 33(12): 2341--2353.

\bibitem[{Hertz et~al.(2019)Hertz, Fogel, Hanocka, Giryes, and
  Cohen-Or}]{hertz2019blind}
Hertz, A.; Fogel, S.; Hanocka, R.; Giryes, R.; and Cohen-Or, D. 2019.
\newblock Blind visual motif removal from a single image.
\newblock In \emph{Proceedings of the IEEE/CVF Conference on Computer Vision
  and Pattern Recognition}, 6858--6867.

\bibitem[{Johnson, Alahi, and Fei-Fei(2016)}]{johnson2016perceptual}
Johnson, J.; Alahi, A.; and Fei-Fei, L. 2016.
\newblock Perceptual losses for real-time style transfer and super-resolution.
\newblock In \emph{Computer Vision--ECCV 2016: 14th European Conference,
  Amsterdam, The Netherlands, October 11-14, 2016, Proceedings, Part II 14},
  694--711. Springer.

\bibitem[{Liang et~al.(2021)Liang, Niu, Guo, Long, and
  Zhang}]{liang2021visible}
Liang, J.; Niu, L.; Guo, F.; Long, T.; and Zhang, L. 2021.
\newblock Visible watermark removal via self-calibrated localization and
  background refinement.
\newblock In \emph{Proceedings of the 29th ACM International Conference on
  Multimedia}, 4426--4434.

\bibitem[{Liu, Zhu, and Bai(2021)}]{liu2021wdnet}
Liu, Y.; Zhu, Z.; and Bai, X. 2021.
\newblock Wdnet: Watermark-decomposition network for visible watermark removal.
\newblock In \emph{Proceedings of the IEEE/CVF Winter Conference on
  Applications of Computer Vision}, 3685--3693.

\bibitem[{Loshchilov and Hutter(2017)}]{loshchilov2017decoupled}
Loshchilov, I.; and Hutter, F. 2017.
\newblock Decoupled weight decay regularization.
\newblock \emph{arXiv preprint arXiv:1711.05101}.

\bibitem[{Park, Tai, and Kweon(2012)}]{park2012identigram}
Park, J.; Tai, Y.-W.; and Kweon, I.~S. 2012.
\newblock Identigram/watermark removal using cross-channel correlation.
\newblock In \emph{2012 IEEE Conference on Computer Vision and Pattern
  Recognition}, 446--453. IEEE.

\bibitem[{Paszke et~al.(2019)Paszke, Gross, Massa, Lerer, Bradbury, Chanan,
  Killeen, Lin, Gimelshein, Antiga et~al.}]{paszke2019pytorch}
Paszke, A.; Gross, S.; Massa, F.; Lerer, A.; Bradbury, J.; Chanan, G.; Killeen,
  T.; Lin, Z.; Gimelshein, N.; Antiga, L.; et~al. 2019.
\newblock Pytorch: An imperative style, high-performance deep learning library.
\newblock \emph{Advances in neural information processing systems}, 32.

\bibitem[{Pei and Zeng(2006)}]{pei2006novel}
Pei, S.-C.; and Zeng, Y.-C. 2006.
\newblock A novel image recovery algorithm for visible watermarked images.
\newblock \emph{IEEE Transactions on information forensics and security}, 1(4):
  543--550.

\bibitem[{Ren et~al.(2019)Ren, Zuo, Hu, Zhu, and Meng}]{ren2019progressive}
Ren, D.; Zuo, W.; Hu, Q.; Zhu, P.; and Meng, D. 2019.
\newblock Progressive image deraining networks: A better and simpler baseline.
\newblock In \emph{Proceedings of the IEEE/CVF conference on computer vision
  and pattern recognition}, 3937--3946.

\bibitem[{Ronneberger, Fischer, and Brox(2015)}]{ronneberger2015u}
Ronneberger, O.; Fischer, P.; and Brox, T. 2015.
\newblock U-net: Convolutional networks for biomedical image segmentation.
\newblock In \emph{Medical Image Computing and Computer-Assisted
  Intervention--MICCAI 2015: 18th International Conference, Munich, Germany,
  October 5-9, 2015, Proceedings, Part III 18}, 234--241. Springer.

\bibitem[{Simonyan and Zisserman(2014)}]{simonyan2014very}
Simonyan, K.; and Zisserman, A. 2014.
\newblock Very deep convolutional networks for large-scale image recognition.
\newblock \emph{arXiv preprint arXiv:1409.1556}.

\bibitem[{Sun, Su, and Wu(2023)}]{sun2023denet}
Sun, R.; Su, Y.; and Wu, Q. 2023.
\newblock DENet: Disentangled Embedding Network for Visible Watermark Removal.
\newblock In \emph{Proceedings of the AAAI Conference on Artificial
  Intelligence}, volume~37, 2411--2419.

\bibitem[{Wang et~al.(2004)Wang, Bovik, Sheikh, and Simoncelli}]{wang2004image}
Wang, Z.; Bovik, A.~C.; Sheikh, H.~R.; and Simoncelli, E.~P. 2004.
\newblock Image quality assessment: from error visibility to structural
  similarity.
\newblock \emph{IEEE transactions on image processing}, 13(4): 600--612.

\bibitem[{Zamir et~al.(2022)Zamir, Arora, Khan, Hayat, Khan, and
  Yang}]{zamir2022restormer}
Zamir, S.~W.; Arora, A.; Khan, S.; Hayat, M.; Khan, F.~S.; and Yang, M.-H.
  2022.
\newblock Restormer: Efficient transformer for high-resolution image
  restoration.
\newblock In \emph{Proceedings of the IEEE/CVF conference on computer vision
  and pattern recognition}, 5728--5739.

\bibitem[{Zhang et~al.(2018)Zhang, Isola, Efros, Shechtman, and
  Wang}]{zhang2018unreasonable}
Zhang, R.; Isola, P.; Efros, A.~A.; Shechtman, E.; and Wang, O. 2018.
\newblock The unreasonable effectiveness of deep features as a perceptual
  metric.
\newblock In \emph{Proceedings of the IEEE conference on computer vision and
  pattern recognition}, 586--595.

\bibitem[{Zhang, Ng, and Chen(2018)}]{zhang2018single}
Zhang, X.; Ng, R.; and Chen, Q. 2018.
\newblock Single image reflection separation with perceptual losses.
\newblock In \emph{Proceedings of the IEEE conference on computer vision and
  pattern recognition}, 4786--4794.

\bibitem[{Zhao, Niu, and Zhang(2022)}]{zhao2022visible}
Zhao, X.; Niu, L.; and Zhang, L. 2022.
\newblock Visible Watermark Removal with Dynamic Kernel and Semantic-aware
  Propagation.

\bibitem[{Zhou et~al.(2018)Zhou, Rahman~Siddiquee, Tajbakhsh, and
  Liang}]{zhou2018unet++}
Zhou, Z.; Rahman~Siddiquee, M.~M.; Tajbakhsh, N.; and Liang, J. 2018.
\newblock Unet++: A nested u-net architecture for medical image segmentation.
\newblock In \emph{Deep Learning in Medical Image Analysis and Multimodal
  Learning for Clinical Decision Support: 4th International Workshop, DLMIA
  2018, and 8th International Workshop, ML-CDS 2018, Held in Conjunction with
  MICCAI 2018, Granada, Spain, September 20, 2018, Proceedings 4}, 3--11.
  Springer.

\end{thebibliography}

\end{document}


\maketitle

\subsection{Effectiveness of Implicit Information Bridge}
In the field of watermark removal, it's common to employ two separate branches for watermark localization and image restoration tasks. To enhance the information flow between these branches, ~\citet{liang2021visible} introduced the MBE module, which utilizes pseudo-masks from the watermark branch to aid the image restoration task in the background branch. Similar designs are also found in ~\cite{cun2021split, zhao2022visible}, where explicit bridges are used to connect the two branches. However, we recognize that this approach might not fully leverage the complete information from both branches.

In fact, watermark localization and image restoration are inherently correlated tasks. Watermark localization naturally implies the areas requiring restoration, while the artifacts discovered during image restoration also contain watermark-related information. Thus, segregating these closely related tasks into separate decoder branches is suboptimal. Consequently, we introduce the concept of implicit branches and employ the Gated-Dconv Feed-Forward Network(GDFN) as an implicit information bridge to facilitate the fusion of these two branches. To validate the rationale behind our design, we conduct comparative experiments by introducing GDFN into the decoding stage of the UNet, as shown in Tab.~\ref{tableA}.

\begin{table}[H]
\centering
\renewcommand\arraystretch{1.5}
\scalebox{0.7}{
\begin{tabular}{c|cccc|cc} 
\toprule\hline
  & PSNR$\uparrow$ & SSIM$\uparrow$ & RMSE$\downarrow$ & RMSEw$\downarrow$ & IoU(\%)$\uparrow$ & F1$\uparrow$ \\
\midrule
w/o GDFN & 41.64 & 0.9928 & 2.96 & 17.51 & 65.71 & 0.7526\\
w/ GDFN & \textbf{41.99} & \textbf{0.9930} & \textbf{2.88} & \textbf{16.93} & \textbf{67.16} & \textbf{0.7657} \\
\bottomrule
\end{tabular}}
\caption{Effects of implicit bridge on LOGO-L.}
\label{tableA}
\end{table}

Here, the UNets we utilized are re-implemented, following the same training approach as WMFormer++. In both cases, a single branch concurrently handles watermark localization and image restoration. To ensure a fair comparison, the parameters of both models are adjusted to achieve comparable levels. Notably, the network using the implicit information bridge demonstrates a more pronounced enhancement in both background image and watermark mask evaluation metrics compared to the network without the implicit information bridge.
This observation strongly attests to the effectiveness of the implicit bridge. It successfully integrates information from both branches, facilitating their meaningful interaction and synergy, thereby substantiating its role in promoting better performance.

\subsection{More Visualizations}
\begin{figure*}[t]
\centering
\includegraphics[width=1\textwidth]{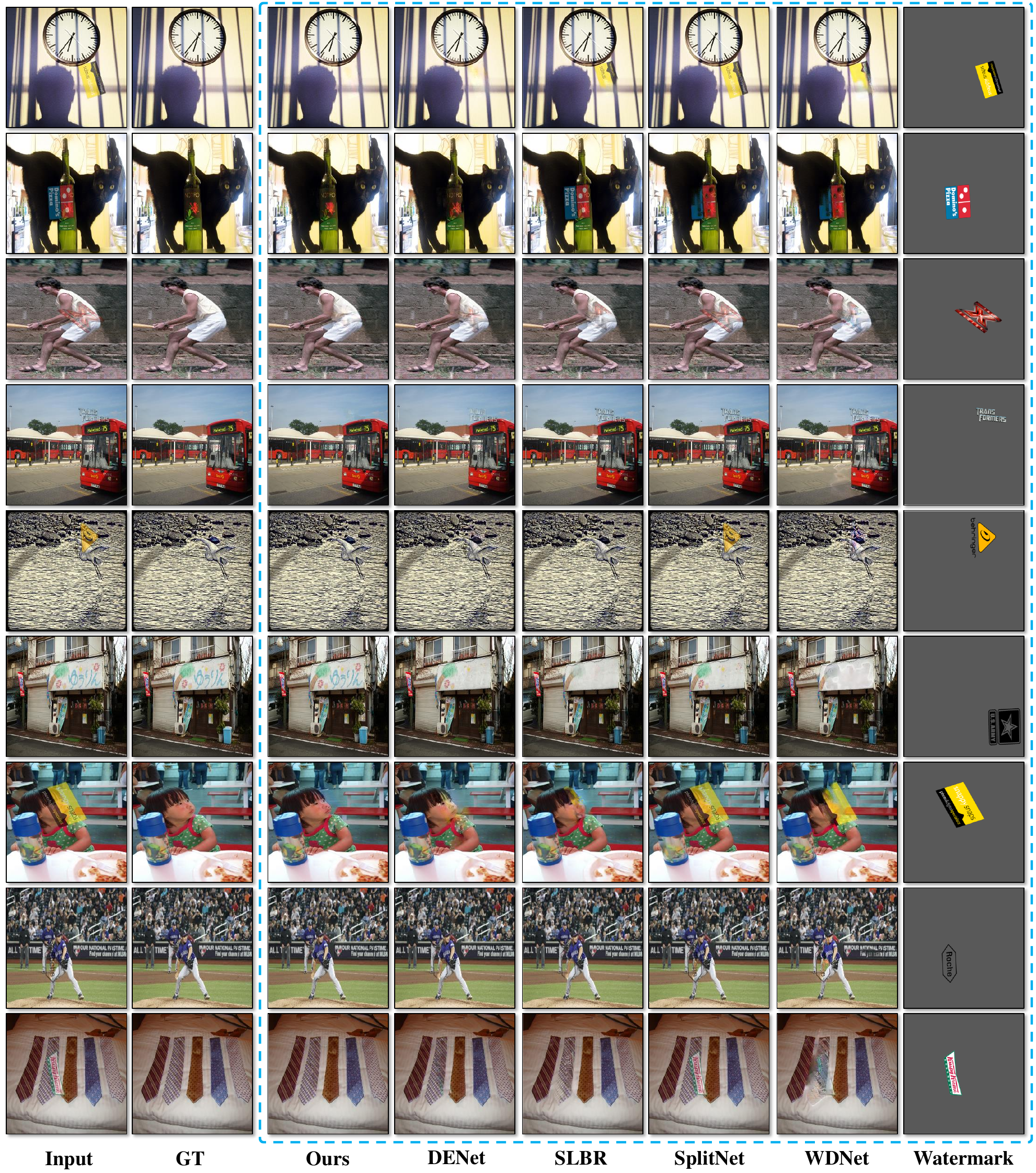} 
\caption{More qualitative comparison of background image restored by different methods on LOGO-L dataset.}
\label{supplement1}
\end{figure*}

\begin{figure*}[t]
\centering
\includegraphics[width=1\textwidth]{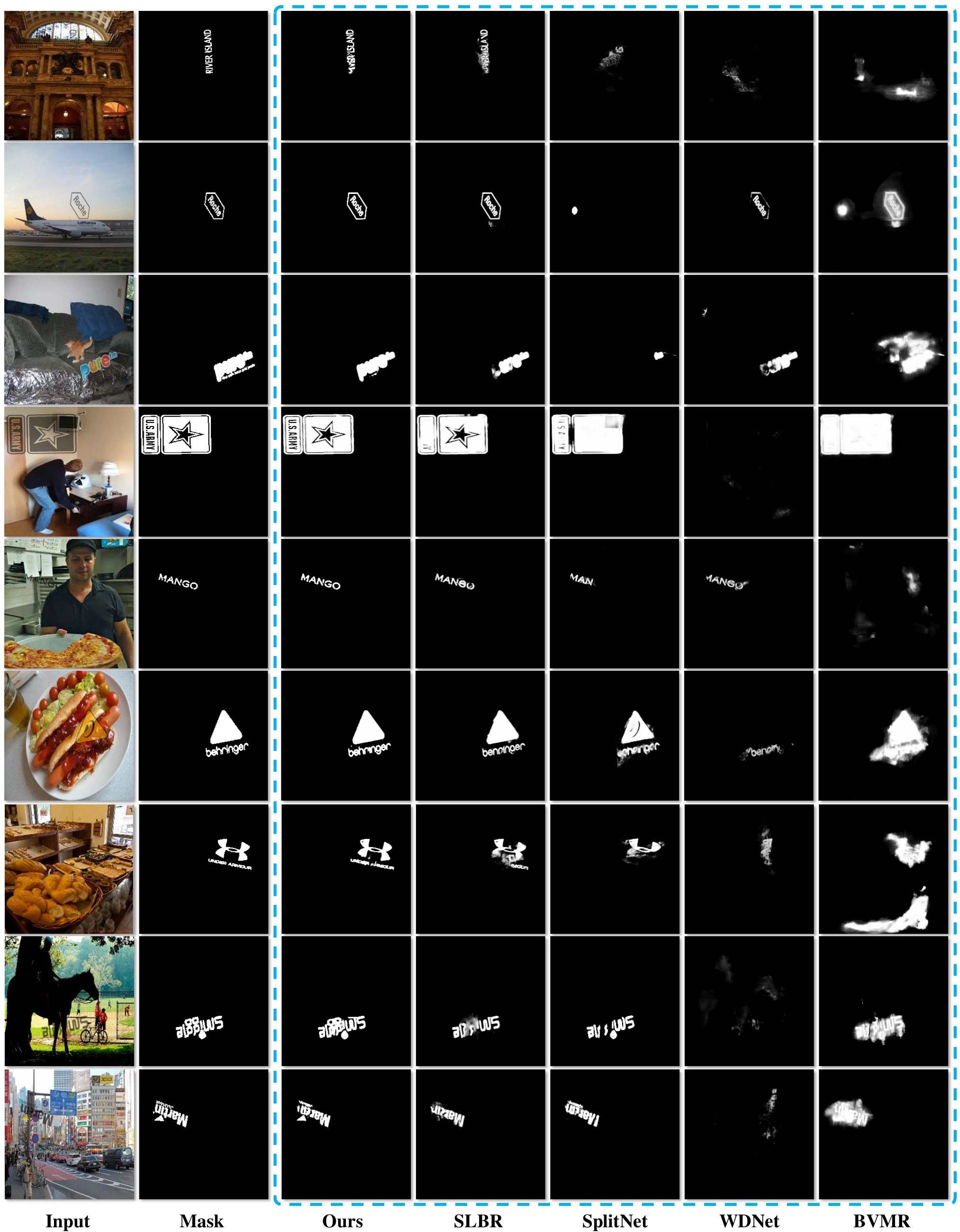} 
\caption{More qualitative comparison of watermark masks located by different methods on LOGO-L dataset.}
\label{supplement2}
\end{figure*}

Similar to Fig.4 and Fig.5 in the main text, we present comparative results against previous state-of-the-art methods\cite{sun2023denet,liang2021visible,cun2021split,liu2021wdnet} to showcase the superiority of our proposed approach. We provide additional visualizations of backgrounds and masks to offer a comprehensive understanding, as depicted in Fig.~\ref{supplement1} and Fig.~\ref{supplement2}.

\begin{figure*}[t]
\centering
\includegraphics[width=0.9\textwidth]{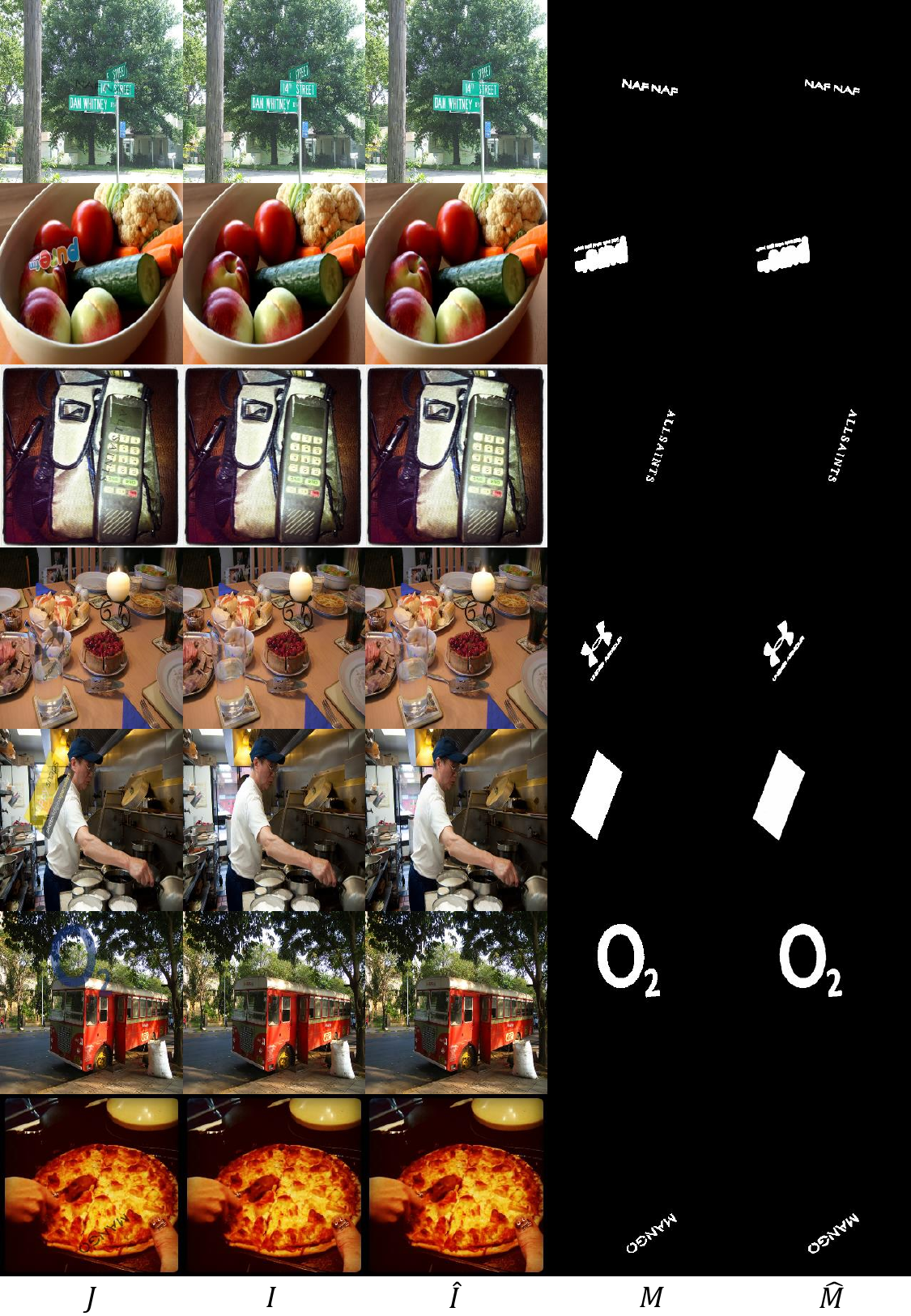} 
\caption{More visualization experiments, where $J$ represents the watermarked image, $I$ represents the ground truth background image, $\hat{I}$ represents the background image predicted by WMFormer++, $M$ represents the ground truth watermark mask, and  $\hat{M}$ represents the watermark mask predicted by WMFormer++.}
\label{supplement3}
\end{figure*}

Furthermore, to thoroughly investigate the output quality of our model, we conduct an extensive comparison between the outcomes produced by our model and the ground truth data, as illustrated in Fig.~\ref{supplement3}.
This in-depth assessment enables a deeper understanding of our method's effectiveness.
\\ \hspace*{\fill} \\
\noindent \textbf{Comparison of restored background image.}
Refer to Fig.~\ref{supplement1}. Overall, our approach exhibits minimal artifacts in the restored background. In the example shown in the sixth row, it is evident that only our method accurately identifies the areas requiring restoration, while other approaches mistakenly interpret the shop's name as the watermark. Moreover, in the last row, our method successfully reconstructs the intricate structural details, which remain challenging for the other methods.
\\ \hspace*{\fill} \\
\noindent \textbf{Comparison of located watermark mask.}
Comparison experiments for watermark localization are presented in Fig.~\ref{supplement2}. Our method demonstrates the most comprehensive and accurate watermark localization. Even in complex scenarios and for watermarks that are difficult for the naked eye to discern, our approach accurately estimates the watermark's position, thanks to the global attention mechanism, as evident in the last two rows.
\\ \hspace*{\fill} \\
\noindent \textbf{Overall quality.}
Finally, we compare the outputs of WMFormer++ with the ground truth in Fig.~\ref{supplement3}. It is evident that the results produced by our method exhibit astonishing precision. Notably, the reconstructed backgrounds exhibit both realism and clarity, and the located watermarks are precisely and finely delineated.
\FloatBarrier

\bibliography{aaai24}